\documentclass[pre,notitlepage,twocolumn,superscriptaddress,color,showpacs]{revtex4-1}
\usepackage{amsmath,amsfonts,amssymb}
\usepackage[pdftex]{graphicx}
\usepackage{xcolor}

\begin{document}
\title{Coevolutionary dynamics of information spreading and heterophilic link rewiring}
\author{Jeehye Choi}
\affiliation{Research Institute for Nanoscale Science and Technology, Chungbuk National University, Cheongju, Chungbuk 28644, Korea}
\author{Byungjoon Min}
\email{bmin@cbnu.ac.kr}
\affiliation{Research Institute for Nanoscale Science and Technology, Chungbuk National University, Cheongju, Chungbuk 28644, Korea}
\affiliation{Department of Physics, Chungbuk National University, Cheongju, Chungbuk 28644, Korea}
\date{\today}

\date{\today}

\begin{abstract}
In many complex systems, the dynamic processes that take place on a network and the changes 
in the network topology are intertwined. Here, we propose a model of coevolutionary dynamics 
of information spreading which is accompanied with link rewiring to facilitate the propagation 
of information. In our model, nodes possessing information attempt to contact new susceptible nodes 
through the link rewiring while the information spreads on a network. 
Using moment-closure and heterogeneous mean-field approximations, we examine 
both the information spread dynamics and network evolution focusing on 
epidemic size, epidemic threshold, and degree distributions at the steady state.
We found that more frequent heterophilic link rewiring leads to a larger epidemic size
but does not alter the epidemic threshold.
We also observed that link rewiring results in a broader degree 
distribution in the steady state. This study provides an insight into the 
the role of the heterophilic link rewiring in both facilitating information propagation 
and inducing network heterogeneity.
\end{abstract}
\maketitle

\section{Introduction}

Complex networks, which mediate interactions among individuals, play a vital role in shaping 
dynamics between them \cite{Albert2002,book_Newman2}. 
For this reason, many studies have been conducted to understand the 
relationship between the structures of complex networks and the dynamics on these networks \cite{book_Newman2}. 
At the same time, the structure of networks constantly change under 
the influence of the state of the nodes \cite{Dorogovtsev,temporal,masuda_book}. 
Moreover, the topology of networks not only influences the dynamic processes occurring within 
them but also evolves in response to the changes in the states of the elements of 
networks \cite{adaptive_book,holme2006,gross2006,gross2008,vazquez2008}.
This feedback inherent in adaptive systems fosters a coevolutionary dynamic between nodes' 
states and network structures, a phenomenon commonly observed in reality. 
Many studies of coevolving networks attempt to integrate these two facets, focusing on the 
coevolution of network topology and nodal dynamics \cite{holme2006,gross2006,gross2008,vazquez2008}.
Research on coevolving networks spans a wide range of contexts, including coevolving voter 
models \cite{vazquez2008,sudo2013,bmin2016,Raducha,bmin2019}, spin systems \cite{Biely2009,Raducha2018,Korbel2023}, 
adaptive epidemic models \cite{gross2006,shaw2008,marceau2010,achterberg2020,achterberg2022}, 
complex contagion models \cite{bmin2023}, game theoretical 
models \cite{Ebel2002,Zimmermann2004}, Boolean network dynamics \cite{Liu2006}, 
and ecological evolution \cite{Dieckmann1999,Drossel2001}.

Epidemic models on coevolving networks offer a powerful means to understand 
the interplay between network topology and disease transmission \cite{adaptive_book,gross2006,gross2008}. 
Traditionally, many studies have investigated the spread of infectious diseases 
through compartmental epidemic models on static networks \cite{pastor,Pastor2001a,Newman2002,choi2022}.
However, real-world networks are often adaptive, meaning that the connections 
between individuals change over time in response to the 
states of their neighbors \cite{adaptive_book}. 
For instance, if an agent identifies an infectious neighbor, that 
node can reduce its interaction with the infectious one to avoid infection \cite{gross2006}. 
To model this scenario, an epidemic model that suppresses epidemic spread 
through link rewiring was proposed \cite{gross2008,marceau2010,Shaw2010}. 
By considering the coevolution of epidemic spread and network structure, 
it might provide insights into effective interventions to mitigate the spread 
of infectious diseases.

Classical epidemic models can represent not only the spread of diseases but also the 
propagation of information, fads, or opinions \cite{pastor,funk2009,kook2021,diaz}. 
In disease transmission, the primary focus is often on preventing the spread of the 
disease \cite{Havlin2003,Fraser2004,Azizi2020}. 
However, in the case of information propagation, the focus is on efficiently delivering 
information to others. Understanding this distinction, instead of employing tactics 
to hinder propagation, we suggest a strategy that amplify the spread of information via link rewiring. 
This change leads us to introduce a coevolutionary dynamics model with link rewiring to 
facilitate information dissemination.
The scenario we are considering is as follows: individuals possessing certain information or 
opinions are motivated to spread that information to a larger fraction of the population.
Since the total amount of interactions with other agents might be limited, it is advantageous for 
information propagation to have more frequent interactions with those who are unaware of the information.
In this context, we propose a model in which individuals break ties with those who already possess 
the information and, instead, establish new links with individuals who do not have the information,
so called ``heterophilic'' link rewiring.

In our study, we derive and analyze moment-closure approximations \cite{keeling2005,gross2006} 
and heterogeneous mean-field approximations \cite{Pastor2001a,marceau2010} to explore the coevolutionary 
dynamics of information propagation. 
Through this analysis, we show the impact of the heterophilic link rewiring 
on the spreading dynamics and the evolution of network structures.
We found that the epidemic size increases with the increase of the rate of link rewiring, while 
the epidemic threshold remains insensitive to the process of link rewiring. 
This suggests that as the network undergoes more frequent rewiring of its connections 
due to the coevolutionary process, the overall size of the epidemic tends to increase. 
In addition, we observed a broadening of the degree distribution during the coevolutionary 
dynamics. We also confirmed that our theoretical frameworks accurately predict all numerical results.

\section{Model}

We propose a model of coevolutionary dynamics based on the susceptible-infectious-susceptible (SIS) 
framework \cite{Pastor2001a,pastor}. In this model, the structures of networks change to promote 
information propagation via 
link rewiring while information spreads into population by following the SIS model. 
While our model involves information spreading, we retain the terminology of the spread 
of infectious disease to maintain consistency with previous studies \cite{pastor}. 
In our model, nodes can exist in either a susceptible or infectious state. 
Here, ``susceptible'' corresponds to a state where nodes do not possess the information. 
Conversely, ``infectious'' corresponds to a state where nodes have the information.

Our model consists of three dynamic processes: infection, recovery, and rewiring. 
Infection and recovery follow the same procedures as in a typical SIS model on networks. 
Specifically, at each time, infectious nodes infect susceptible neighbors at 
a rate $p$. Each infectious node also has a rate $r$ of autonomously returning to 
a susceptible state, representing the recovery process. In addition to infection 
and recovery, our model introduces a link rewiring process that forms the basis 
of the coevolutionary dynamics. At each time, links connecting two infectious 
nodes are removed at a rate $w$. Subsequently, one infectious node out of two nodes that 
used to form a link seeks a new neighbor among the susceptible nodes and forms a new connection. 
The heterophilic link rewiring process allows the network structure to dynamically respond 
to the spread of information, leading to an interplay between information dissemination 
and changes in network topology.

\begin{figure}[t]
\includegraphics[width=0.9\columnwidth]{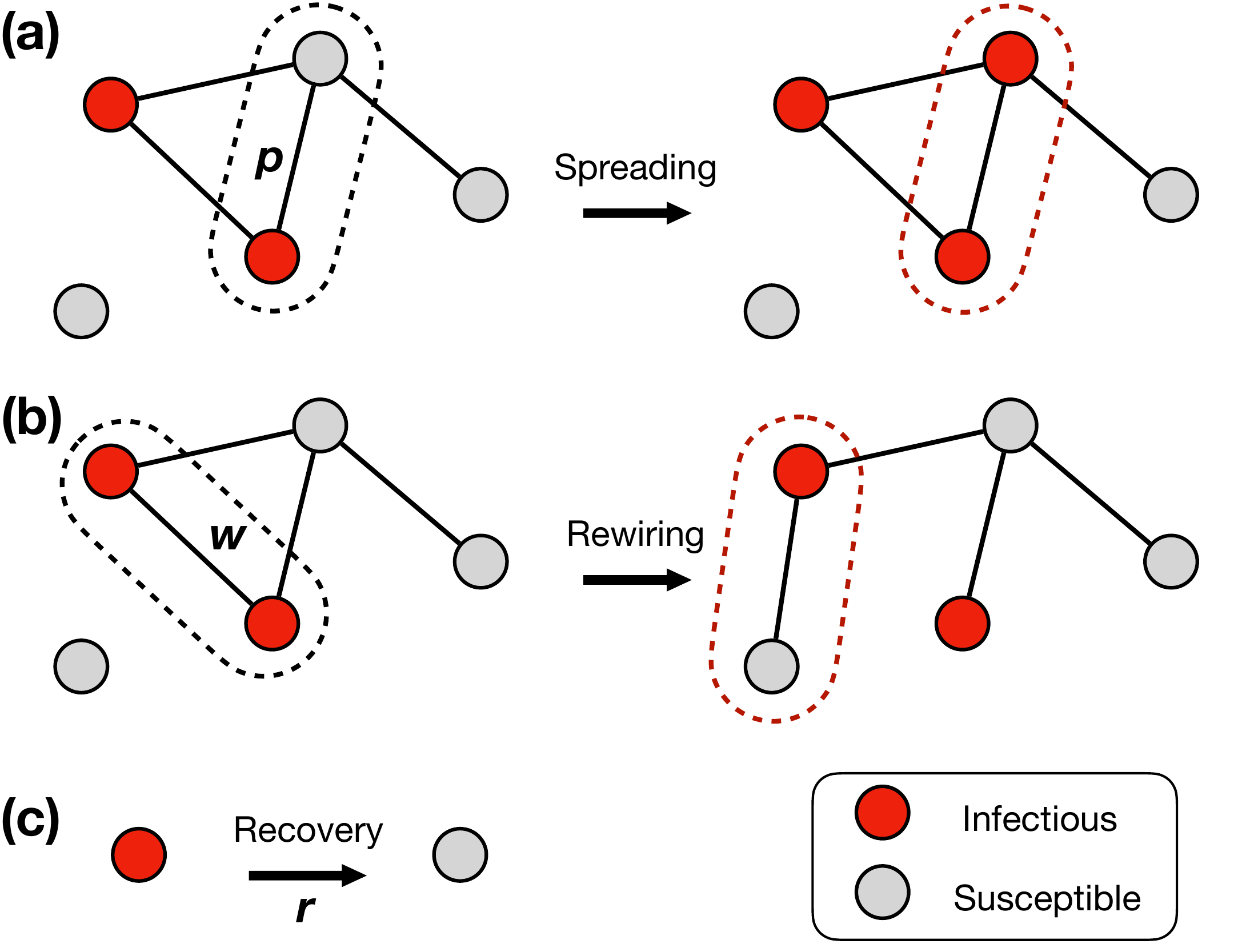}
\caption{
Our coevolutionary model consists of three dynamical processes: 
(a) information spreading occurring at a rate of $p$, through links connecting susceptible 
and infectious nodes; (b) link rewiring at a rate of $w$, which involves removing 
a link between a pair of infectious nodes and establishing a new link between 
infectious and susceptible nodes; and (c) recovery at a rate of $r$, which an infectious 
node autonomously transitions to a susceptible state.
}
\label{fig1}
\end{figure}

\section{Theory}

\subsection{Moment-closure approximation}

To investigate the coevolutionary dynamics, we first attempt to derive 
the time evolution of the densities of susceptible and infectious nodes, denoted 
as $S$ and $I$, respectively. By definition, they satisfy the normalization condition 
$S+I=1$. When we define the fraction of $\textrm{SI}$ links $N$ as $\ell_\textrm{SI}$, the 
evolution of $I$ can be expressed as
\begin{align}
\frac{dI}{dt} =p \ell_\textrm{SI}-r I,
\label{eq:di}
\end{align}
where $p$ and $r$ represent the rates of infection and recovery, respectively. 
As shown in Eq.~(\eqref{eq:di}), the fraction of the link $\textrm{SI}$ is required to predict 
the evolution of both $I$ and $S$.

We further define the fraction of $\textrm{SS}$ and $\textrm{II}$ links 
to the number of nodes $N$ as $\ell_\textrm{SS}$ and $\ell_\textrm{II}$. 
The sum of $\ell_\textrm{II}$, $\ell_\textrm{SS}$, and $\ell_\textrm{SI}$ satisfies 
the following condition: $\ell_\textrm{SS}+ \ell_\textrm{II}+ \ell_\textrm{SI}=\langle k \rangle / 2$, where 
$\langle k \rangle$ represents the average degree. 
The time evolution of link densities involves the densities of the triplets, for example, the 
densities of $\ell_\textrm{ISI}$, $\ell_\textrm{IIS}$, $\ell_\textrm{SSS}$, etc, because
$i$-th moments will depend on the $(i+1)$-th moments.
Similarly, higher-order moments are recursively needed in principle, up to infinite order terms.

Here, we employ the moment-closure approximation \cite{keeling2005,gross2006} to terminate 
the recursion. In the following, we explain how to decompose the density of triplets into link 
densities based on the moment-closure approximations. We begin by approximating the density of 
triplet $\ell_\textrm{ABC}$, where $\textrm{A}$, $\textrm{B}$, and $\textrm{C}$ represent 
the states of nodes (e.g., $\textrm{S}$ and $\textrm{I}$ in our example). One half of the 
$\textrm{ABC}$-triplet is an AB link, which occurs at density $\ell_\textrm{AB}$. 
The other half is an  $\textrm{BC}$ link that share $\textrm{B}$ node with an  $\textrm{AB}$ link, that is $\textrm{ABC}$-triplets.
To approximate the number of $\textrm{ABC}$-triplets, we estimate the number of nodes with state $\textrm{C}$ 
that are connected to node with state $\textrm{B}$ in $\textrm{AB}$ links. 
Since we arrived at the node with state $\textrm{B}$ by following a link, the distribution of the expected 
number of neighbors of that node can be approximated as $(k-1) P(k)/\langle k \rangle$
where $P(k)$ is the degree distribution of a network, assuming the networks are uncorrelated.
Each of these links is a BC-link with probability $\ell_\textrm{BC}/(\langle k \rangle {B})$. 
Then the estimated number of nodes with state $\textrm{C}$ that are connected to node with state 
$\textrm{B}$ in $\textrm{AB}$ links is 
$\frac{\langle k^2 \rangle - \langle k \rangle}{\langle k \rangle^2} \frac{\ell_\textrm{BC}}{B}$

In general, $\langle k^2 \rangle$ at a steady state is not known a priori because the degree 
distribution varies by link rewiring. 
In this study, we use Erd\H{o}s-R\'enyi (ER) graphs as the initial network configuration. 
On ER graphs, $\frac{\langle k^2 \rangle - \langle k \rangle}{\langle k \rangle^2}$ is 
unity in the thermodynamic limit. However, the value changes in time since the rewiring in our model 
is not entirely random. If we neglect the heterogeneity of the degree distribution 
caused by the link rewiring, we can assume that this value is approximately $1$ regardless of time. 
Taking the density of $\textrm{AB}$-link and the probability that they connect to 
an additional $\textrm{BC}$-link into account, we obtain an approximation for the density of triples as 
$\ell_\textrm{ABC} \approx \ell_\textrm{AB}\ell_\textrm{BC}/B$ \cite{keeling2005,gross2006}.

Based on the approximation, we can derive the time evolution of link densities:
\begin{align}
\frac{d \ell_\textrm{II}}{dt} &= p \ell_\textrm{SI} \left( \frac{\ell_\textrm{SI}}{S}+1\right)-(2r+w)\ell_\textrm{II},\\
\frac{d \ell_\textrm{SS}}{dt} &= r \ell_\textrm{SI}- \frac{2p \ell_\textrm{SI} \ell_\textrm{SS}}{S}.
\label{eq:dl}
\end{align}
The normalization condition such that $\ell_\textrm{SS}+ \ell_\textrm{II}+ \ell_\textrm{SI}=\langle k \rangle / 2$
gives the time evolution of $\ell_\textrm{SI}$. 
The trivial solution of the coupled equations given in Eqs.~(\ref{eq:di}-\ref{eq:dl}) at the steady state
is $(I, \ell_\textrm{SI}, \ell_\textrm{SS})=(0,0,1)$, corresponding to the disease-free phase. 
In this phase, all nodes are in the susceptible state, so that all links are also $\textrm{SS}$ state.
The non-trivial solution at the steady state of the equations can be computed as
\begin{widetext}
\begin{align}
I&=\frac{1}{r (2 p + w)} \left[ p (r + K r + K w) - \sqrt{2 r^3 (2 p + w) + [r w + p (r - 2 K r - K w)]^2} \right], \\
\ell_\textrm{SI}&=\frac{1}{p (2 p + w)} \left[ p (r + 2 K r + K w) - \sqrt{2 r^3 (2 p + w) + [r w + p (r - 2 K r - K w)]^2} \right].
\end{align}
\end{widetext}
The non-trivial solution corresponds to the active epidemic phase with a non-zero value of $I$.

The onset of the epidemic phase can be predicted by the linear stability analysis of the Jacobian 
matrix of Eqs.~(\ref{eq:di}-\ref{eq:dl}). The epidemic threshold is located at where the largest 
eigenvalue of the Jacobian matrix at $(I, \ell_\textrm{SI}, \ell_\textrm{SS})=(0,0,1)$ becomes
zero. When we apply the linear stability analysis, we arrive at the epidemic threshold as
\begin{align}
p_c = \frac{r}{\langle k \rangle},
\label{eq:pc}
\end{align}
regardless of the value of $w$. It implies that the link rewiring process does not change 
the location of the epidemic threshold, but it affects the steady-state density of infectious nodes.

\subsection{Heterogeneous mean-field approximations}

We next derive heterogeneous mean-field (HMF) equations for our model \cite{marceau2010}. 
By using the HMF approximations, we attempt to obtain not only the fraction of infectious nodes
but also the network structure that emerges at the steady state. 
Let $S_{k,q}$ and $I_{k,q}$ respectively be the fraction of susceptible and infectious nodes 
with degree $k$ and the number $q$ of infectious neighbors. 
Using these defined fractions for different states of nodes, we can derive 
the differential equations for the susceptible and infectious fractions, $S_{k,q}$ and 
$I_{k,q}$ \cite{marceau2010},
\begin{widetext}
\begin{align} 
\label{diffeq1}  
\frac{dS_{k,q}}{dt}&=r I_{k,q}-p q S_{k,q}+r \left[(q+1)S_{k,q+1}-qS_{k,q}\right] 
+p \frac{S_\textrm{SI}}{S_\textrm{S}}\left[(k-q+1)S_{k,q-1}-(k-q)S_{k,q} \right] 
+  \frac{w I_\textrm{I}}{2S} \left[ S_{k-1,q-1}-S_{k,l}\right],~\\
\label{diffeq2} 
\frac{dI_{k,q}}{dt}&=-r I_{k,q}+p q S_{k,q} +r \left[(q+1)I_{k,l+1}- q I_{k,l}\right]  
+ p \left(1+\frac{S_\textrm{II}}{S_\textrm{I}}\right)\left[(k-q+1)I_{k,q-1}-(k-q)I_{k,q} \right] \\
&\quad +w\left[\frac{1}{2}(q+1)I_{k+1,q+1}+\frac{1}{2}(q+1)I_{k,q+1} - qI_{k,q}\right], \nonumber
\end{align}
\end{widetext}
where the first moments are defined as
\begin{align}
S_\textrm{S}=&\sum_{k,q} (k-q)S_{k,q},\quad 
S_\textrm{I}=\sum_{k,q} q S_{k,q},\\
I_\textrm{S}=&\sum_{k,q} (k-q) S_{k,q}, \quad 
I_\textrm{I}=\sum_{k,q} q I_{k,q},
\label{momt1}
\end{align}
and the second moments are defined as
\begin{align}
S_\textrm{SI}=\sum_{k,q} q (k-q) S_{k,q}, \quad
S_\textrm{II}=\sum_{k,q} q (q-1) S_{k,q}.
\label{momt2}
\end{align}

The meaning of each term in Eqs.~(\eqref{diffeq1} and \eqref{diffeq2}) is transparent.
The first term of Eq.~(\eqref{diffeq1}), $r I_{k,q}$, represents the recovery 
process from infectious nodes to susceptible nodes. The second term represents 
the infection process of susceptible nodes through infectious neighbors $q$ 
with the infection rate $p$. The third term represents the recovery of infectious
neighbors' for susceptible nodes, meaning that $S_{k,q} \rightarrow S_{k,q-1}$.
The fourth term represents the infection of susceptible neighbors' for susceptible 
nodes, meaning that $S_{k,q} \rightarrow S_{k,q+1}$.
The last term stands for the susceptible nodes which are obtained a new 
connection through rewiring. Each term in Eq.~(\eqref{diffeq2}) can be explained similarly. 
Note that the last term in Eq.~(\eqref{diffeq2}) corresponds to the breaking of 
$\textrm{II}$ links and rewiring of $\textrm{SI}$ links.

To calculate the density of susceptible and infectious nodes at the steady state, we 
solve Eqs. (\ref{diffeq1},\ref{diffeq2}) by numerical iterations.
We can determine the density of susceptible ($S$) and infectious ($I$) nodes by summing 
$S_{k,q}$ and $I_{k,q}$ over all possible values of $k,q$, respectively. 
That is, the densities of $S$ and $I$ nodes are given by
\begin{align}
S=\sum_{k,q} S_{k,q},\quad
I=\sum_{k,q} I_{k,q}.
\end{align}

Additionally, we can obtain the degree distributions for susceptible $P_S(k)$
and infectious $P_I(k)$ nodes at the steady state. These distributions represent the probability 
of finding a node with a degree $k$ in the susceptible or infectious state, respectively. 
We can calculate the degree distributions for susceptible and infectious nodes by summing 
$S_{k,q}$ and $I_{k,q}$ over $q$, then dividing by the total density of $S$ and $I$ nodes, respectively. 
The equations for calculating the degree distributions of infectious and susceptible nodes are 
given by 
\begin{align}
P_\textrm{I}(k) = \frac{1}{I} \sum_{q=0}^{k} I_{k,q}, \quad 
P_\textrm{S}(k) = \frac{1}{S} \sum_{q=0}^{k} S_{k,q}.
\end{align}

\begin{figure}[t]
\includegraphics[width=\columnwidth]{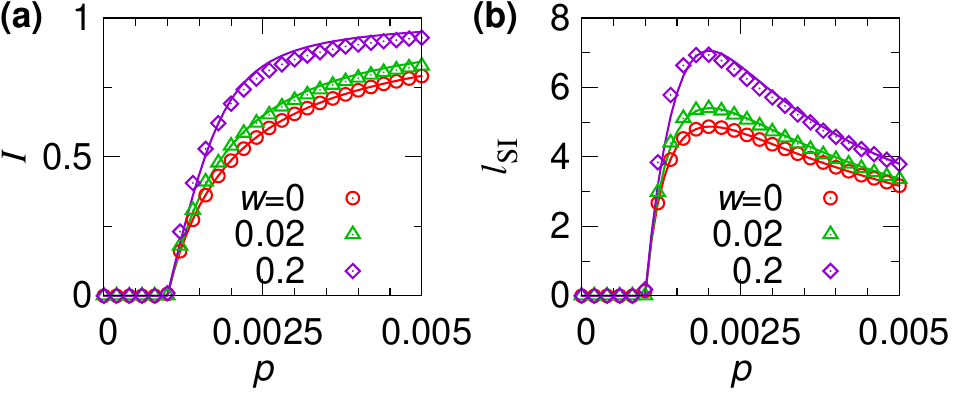}
\caption{The density of (a) infectious nodes and (b) the fraction of
SI links for each node with respect to the infection rate $p$,
with various values of rewiring rate $w$ are shown.
The symbols represent numerical simulations obtained with $N=10^5$, $\langle k \rangle=20$, and $r=0.02$.
The lines represent theoretical predictions based on Eqs.~(4,5).
}
\label{fig2}
\end{figure}

\section{Results}

We investigate the role of the link rewiring in the propagation of information 
in coevolutionary dynamics. We explore the density $I$ of infectious 
nodes, which represents the extent of information spread. 
Figure 2(a) shows the density of infectious nodes as a function of infection rate $p$
for various values of link rewiring rates, $w=0,0.02,0.2$. 
We use ER graphs as the initial structure of networks with a size of $N=10^5$ and 
mean degree $\langle k \rangle=20$. 
Our theoretical predictions based on moment-closure approximations show a great 
agreement with the numerical simulations across different parameter values.

Our results reveal a direct relationship between the size of the epidemic, representing the 
extent of information spread, and the link rewiring rate $w$. 
As the link rewiring rate $w$ increases, the size of the epidemic increases correspondingly. 
This finding indicates that the more frequent rewiring of links by infectious nodes 
in search of susceptible neighbors enhances the spread of information. 
However, the epidemic threshold remains constant with respect to $w$, as predicted by 
Eq.~(\eqref{eq:pc}). It implies that the heterophilic link rewiring has no impact on coevolutionary 
dynamics when $\textrm{II}$ links do not extensively exist, leading to the result where $p_c$ is 
independent of $w$. However, as $p$ exceeds $p_c$, rewiring plays a role in increasing 
the fraction of infectious nodes.

\begin{figure}[t]
\includegraphics[width=\columnwidth]{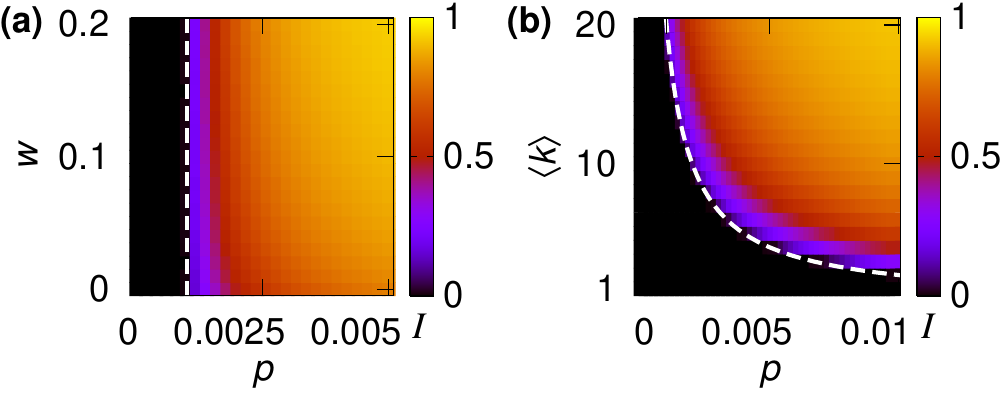}
\caption{
The fraction $I$ of infectious nodes with respect to (a) $w$ and $p$
with the mean degree $\langle k \rangle = 20$ and 
(b) $\langle k \rangle$ and $p$ with $w=0.02$ are shown.
The dashed lines correspond to the epidemic threshold predicted by Eq.~(\eqref{eq:pc}).
The numerical results are obtained with $N=10^5$ and $r=0.02$.
}
\label{fig3}
\end{figure}

We next examine the average fraction $\ell_\textrm{SI}$ of SI links, indicating the density 
of active interactions for information spreading between susceptible and infectious nodes. 
Figure 2(b) illustrates the behavior of the average fraction  $\ell_\textrm{SI}$ of SI links. 
The lines in this figure represent the values of $\ell_\textrm{SI}$ obtained 
from our moment-closure approximations, as given in Eq.~(5). The fraction $\ell_\textrm{SI}$ 
remains null below $p_c$ because there are no infectious nodes, but begins to increase at $p_c$. 
It exhibits a sharp increase initially, but then starts to decline as the growth of $I$ slows down. 
In addition, we found that as the link rewiring rate $w$ increases, $\ell_\textrm{SI}$ reaches 
higher values, reflecting the higher density of infectious nodes as $w$ increases.

We show the numerical results of the density $I$ of infectious nodes as a function of $w$ and $p$ 
in Fig.~3(a)and as a function of $\langle k \rangle$ and $p$ in Fig.~3(b). Once again, we found
that the epidemic threshold predicted by Eq.~(5), denoted in dotted lines in Figs.~3(a,b) 
matches well with the numerical simulations. Figure~3(a) shows that as the link rewiring rate $w$
increases, the size of the epidemic increases. Figure~3(b) shows that for various values of $\langle k \rangle$ 
the epidemic threshold follows the relation in Eq.~(6). We also confirm that while 
the extent of information spread is sensitive to link rewiring, the epidemic threshold remains 
unaffected by changes in network topology.

Finally, we study the network structure that emerges at the steady state of the adaptive dynamics. 
We particularly focus on how the rewiring influences the degree distribution in the network. 
Figure~\ref{fig3} exhibits the degree distributions of infectious $P_{\textrm{I}}(k)$ and susceptible nodes $P_{\textrm{S}}(k)$
with various values of $w=0,0.02,0.2$.
The theoretical predictions obtained from the HMF approximations denoted by lines in Fig.~3 
show a great agreement with numerical results.
We found  that a higher rewiring rate results in a more heterogeneous degree distribution at the steady 
state, even though the networks initially share identical structures.
When the rewiring rate increases, the fraction of isolated nodes also increases. 
This is because when an $\textrm{II}$ link is broken, one of the nodes are connected to a susceptible node 
while the other loses its degree. 
This implies that increased rewiring rate in coevolutionary dynamics of information spreading 
can be a source of the heterogeneity of network topology.

\begin{figure}[t]
\includegraphics[width=\columnwidth]{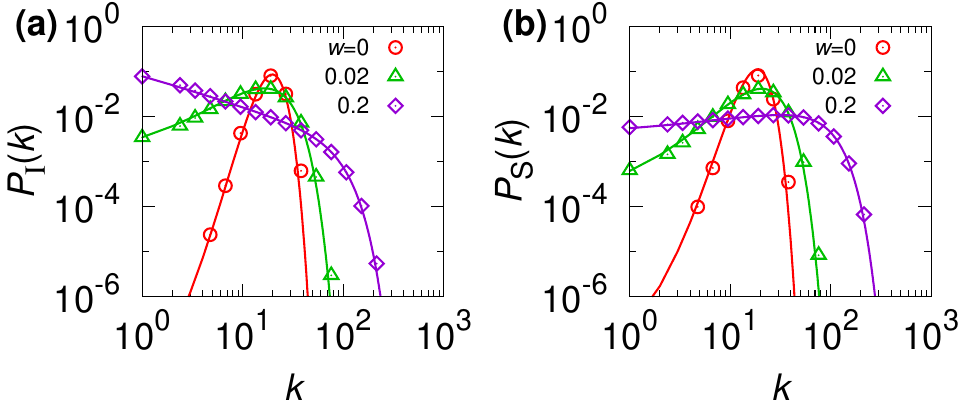}
\caption{
Degree distribution of (a) infectious nodes $P_I(k)$ and (b) susceptible nodes 
at the steady state with various values of $w$. 
The numerical results denoted by symbols are averaged over $500$ realizations with the 
parameters $N=10^5$, $\langle k \rangle=20$, $p=0.005$ and $r=0.02$.
The analytical predictions denoted by lines are obtained by the HMF.
}
\label{fig3}
\end{figure}

\section{Discussion}

In this study, we explore coevolutionary dynamics of information spreading that 
combines network structural changes with spreading dynamics. We use 
moment-closure approximations and heterogeneous mean-field analysis for the scenario 
where link rewiring occurs by infectious nodes in search of susceptible neighbors 
to enhances the spread of information. 
Our research highlights the impact of link rewiring on information spread dynamics, emphasizing 
its effects on epidemic size, epidemic threshold, and network structures.
We found that that the epidemic size increases with the increase of rewiring rate, but the 
epidemic threshold remains insensitive to the process of link rewiring. 
Furthermore, we observe more heterogeneous degree distributions at the steady state
as the link rewiring rate increases. Our study presents a coevolutionary spreading model 
that utilizes link rewiring as a catalyst for both information propagation and a source 
of the heterogeneity in network topology.

\acknowledgments

We thank N. Masuda for helpful discussion.
This work was supported by the National Research Foundation of Korea (NRF) 
grant funded by the Korea government (MSIT) (no.~2020R1I1A3068803).


\begin{thebibliography}{99}
\bibitem{Albert2002}  R. Albert and A.-L. Barabási, Statistical mechanics of complex networks, 
	Rev. Mod. Phys. {\bf 74}, 47–49 (2002).
\bibitem{book_Newman2} M. Newman, A.-L. Barabási and D. J. Watts, The Structure and Dynamics of Networks 
	(Princeton University Press, 2006).
\bibitem{Dorogovtsev} S. N. Dorogovtsev and J. F. F. Mendes, Evolution of networks, 
	Adv. Phys. {\bf 51}, 1079-1187 (2002).
\bibitem{temporal} P. Holme and J. Saram\"aki, Temporal networks, 
	Phys. Rep. {\bf 519}, 97-125 (2012).
\bibitem{masuda_book} N. Masuda and R. Lambiotte,
	A guide to temporal networks (World Scientific, 2020).
\bibitem{holme2006} P. Holme and M. E. J. Newman, 
	Nonequilibrium phase transition in the coevolution of networks and opinions,
	Phys. Rev. E {\bf 74}, 056108 (2006).
\bibitem{adaptive_book} T. Gross and H. Sayama, 
	Adaptive networks: theory, models and applications (Springer, 2009).
\bibitem{gross2006} T. Gross, C. J. D. D'Lima, and B. Blasius, 
	Epidemic dynamics on an adaptive network,
	Phys. Rev. Lett. {\bf 96}, 208701 (2006).
\bibitem{gross2008} T. Gross and B. Blasius, 
	Adaptive coevolutionary networks: a review,
	J. R. Soc. Interface {\bf 5}, 259-271 (2008).
\bibitem{vazquez2008} F. Vazquez, V. M. Egu\'iluz, and M. San Miguel, 
	Generic absorbing transition in coevolution dynamics,
	Phys. Rev. Lett. {\bf 100}, 108702 (2008).
\bibitem{sudo2013} S. D. Yi, S. K. Baek, C.-P. Zhu, and B. J. Kim, 
	Phase transition in a coevolving network of conformist and constrarian voters,
	Phys. Rev. E {\bf 87}, 012806 (2013).
\bibitem{bmin2016} B. Min and M. San Miguel, 
	Fragmentation transitions in a coevolving nonlinear voter model, 
	Sci. Rep. {\bf 7}, 12864 (2017).
\bibitem{Raducha} T. Raducha, B. Min, and M. San Miguel, 
	Coevolving nonlinear voter model with triadic closure, 
	EPL (Europhys. Lett.) {\bf 123}, 30001 (2018).	
\bibitem{bmin2019} B. Min and M. San Miguel, 
	Multilayer coevolution dynamics of the nonlinear voter model, 
	New J. Phys. 21, 035004 (2019).
\bibitem{Biely2009} C. Biely, R. Hanel, and S. Thurner,
	Socio-economical dynamics as a solvable spin system on co-evolving networks, Eur. Phys. J. B {\bf 67}, 285–289 (2009).
\bibitem{Raducha2018} T. Raducha, M. Wilinski, T. Gubiec, and H. E. Stanley, 
	Statistical mechanics of a coevolving spin system, Phys. Rev. E {\bf 98}, 030301 (2018).
\bibitem{Korbel2023} J. Korbel, S. D. Lindner, T. M. Pham, R. Hanel and S. Thurner, 
	Homophily-Based Social Group Formation in a Spin Glass Self-Assembly Framework, Phys. Rev. Lett. {\bf 130}, 057401 (2023).
\bibitem{shaw2008} L. B. Shaw and I. B. Schwartz, Fluctuating epidemics on adaptive networks, 
	Phys. Rev. E {\bf 77}, 066101 (2008).
\bibitem{marceau2010} V. Marceau, P.-A. No\"el, L. H\'ebert-Dufresne, A. Allard, and L. J. Dub\'e,
	Adaptive networks: coevolution of disease and topology, Phys. Rev. E {\bf 82}, 036116 (2010).
\bibitem{achterberg2020} M. A. Achterberg, J. L. A. Dubbeldam, C. J. Stam, and P. Van Mieghem,
	Classification of link-breaking and link-creation updating rules in susceptible-infected-susceptible epidemics on adaptive networks,
	Phys. Rev. E  {\bf 101}, 052302 (2020).
\bibitem{achterberg2022} M. A. Achterberg and P. Van Mieghem,
	Moment closure approximations of susceptible-infected-susceptible epidemics on adaptive networks, 
	Phys. Rev. E {\bf 106}, 014308 (2022).	
\bibitem{bmin2023} B. Min and M. San Miguel, Threshold cascade dynamics on coevolving networks, Entropy {\bf 25}, 929 (2023).
\bibitem{Ebel2002} H. Ebel and S. Bornholdt, Coevolutionary games on networks, 
	Phys. Rev. E {\bf 66}, 056118 (2002).
\bibitem{Zimmermann2004} M. G. Zimmermann, V. M. Egu\'{\i}luz, and M. San Miguel, 
	Coevolution of dynamical states and interactions in dynamic networks, 
	Phys. Rev. E {\bf 69}, 065102 (2004).
\bibitem{Liu2006} M. Liu and K. E. Bassler, 
	Emergent criticality from coevolution in random Boolean networks, Phys. Rev. E {\bf 74}, 041910 (2006).
\bibitem{Dieckmann1999} U. Dieckmann and M. Doebeli, On the origin of species by sympatric speciation, 
	Nature {\bf 400}, 354–357 (1999).
\bibitem{Drossel2001} B. Drossel, P. G. Higgs, and A. J. McKane,
	The influence of predator–prey population dynamics on the long-term evolution of food web structure, 
	J. Theor. Biol. {\bf 208}, 91–107 (2001). 
\bibitem{Pastor2001a} R. Pastor-Satorras and A. Vespignani, 
	Epidemic Spreading in Scale-Free Networks, Phys. Rev. Lett. {\bf 86}, 3200 (2001).	
\bibitem{Newman2002} M. E. J. Newman, Spread of epidemic disease on networks, Phys. Rev. E {\bf 66}, 016128 (2002).
\bibitem{pastor} R. Pastor-Satorras, C. Castellano, P. Van Mieghem, and A. Vespignani,
	Epidemic processes in complex networks, 
	Rev. Mod. Phys. {\bf 87}, 925 (2015).
\bibitem{choi2022} J. Choi and B. Min, Identifying influential subpopulations in metapopulation 
	epidemic models using message-passing theory, 
	Phys. Rev. E {\bf 105}, 044308 (2022).
\bibitem {Shaw2010} L. B. Shaw and I. B. Schwartz, 
	Enhanced vaccine control of epidemics in adaptive networks, 
	Phys. Rev. E {\bf 81}, 046120 (2010).
\bibitem{funk2009} S. Funk, E. Gilad, C. Watkins, and V. A. A. Jansen, 
	The spread of awareness and its impact on epidemic outbreaks, 
	Proc. Nat. Acad. Sci. USA {\bf 106(16)}, 6872-6877 (2009).
\bibitem{kook2021} J. Kook, J. Choi, and B. Min, 
	Double transitions and hysteresis in heterogeneous contagion processes, 
	Phys. Rev. E {\bf 104}, 044306 (2021).
\bibitem{diaz} F. Diaz-Diaz, M. San Miguel and S. Meloni, 
	Echo chambers and information transmission biases in homophilic and heterophilic networks, 
	Sci. Rep. {\bf 12}, 9350 (2022).  
\bibitem{Havlin2003} R. Cohen, S. Havlin, and D. Ben-Avraham, 
	Efficient immunization strategies for computer networks and populations,
	Phys. Rev. Lett. {\bf 91} 247901 (2003).
\bibitem {Fraser2004} C. Fraser, S. Riley, R. M. Anderson, and N. M. Ferguson, 
	Factors that make an infectious disease outbreak controllable, 
	Proc. Natl. Acad. Sci. USA {\bf 101}, 6146-6151 (2004).
\bibitem{Azizi2020} A. Azizi, C. Montalvo, B. Espinoza, Y. Kang, and C. Castillo-Chavez, 
	Epidemics on networks: Reducing disease transmission using health emergency declarations and peer communication, 
	Infect. Dis. Model. {\bf 5}, 12-22 (2020).
\bibitem{keeling2005} M. J. Keeling and K. T. D. Eames, 
	Networks and epidemic models, 
	J. R. Soc. Interface {\bf 2(4)}, 295–307  (2005).

\end{thebibliography}
\end{document}